\documentclass[twocolumn,superscriptaddress,amsmath, amssymb, amsfonts,preprintnumbers,aps,prd,longbibliography,nofootinbib]{revtex4-2}

\renewcommand{\sec}[1]{\textit{#1. --- }}

\usepackage{graphicx}
\usepackage{dcolumn}
\usepackage{bm}
 \usepackage{amstext}
 \usepackage{amssymb}
\usepackage{subcaption}
 \usepackage{amsmath}
 \usepackage{graphicx}
 \usepackage{color}
 \usepackage{bbold}
 \usepackage{delimset} 
\usepackage[colorlinks=true]{hyperref}
\usepackage{orcidlink}
\usepackage{float}

\newcommand{\tabeq}[2]{ \parbox{#1}{  \be\begin{aligned}#2 \end{aligned} \nonumber \ee }}

\newcommand{\CricArrowRight}[1]{%
    \setlength{\@SizeOfCirc}{\maxof{\widthof{#1}}{\heightof{#1}}}%
    \tikz [x=1.0ex,y=1.0ex,line width=.15ex, draw=blue]%
        \draw [->,anchor=center]%
            node (0,0) {#1}%
            (0,1.2\@SizeOfCirc) arc (85:-240:1.2\@SizeOfCirc);%
}%

\usetikzlibrary{decorations.pathmorphing}
\usetikzlibrary{decorations.markings}
\usetikzlibrary{positioning, shapes, snakes, arrows}
 \usepackage{tikz-feynman}

\tikzset{ 
	graviton/.style={line width=.8pt, -latex,decorate, decoration={snake, segment length=4pt,amplitude=1pt, pre length=.1cm, post length=.25cm}},
	worldline/.style={gray, line width=1pt},
	worldlineBold/.style={black, line width=.6pt},
        background/.style={black,dotted,line width=1pt},
	zUndirected/.style={line width=1pt},
	zParticle/.style={line width=1pt,postaction={decorate},decoration={markings,mark=at position .6 with {\arrow[#1]{latex}}}},
	zParticleF/.style={line width=1pt,postaction={decorate}},
	cscalar/.style={line width=1pt,postaction={decorate},decoration={markings,mark=at position .6 with {\arrow[#1]{latex}}}},
	cscalar2/.style={line width=1pt,postaction={decorate},decoration={markings,mark=at position .8 with {\arrow[#1]{latex}}}},
	photon/.style={line width =.8pt, decorate, decoration={snake, segment length=3pt, amplitude=1.2pt,  pre length=.1cm, post length=.1cm}},
	 mid arrow/.style={postaction={decorate,decoration={
        markings,
        mark=at position .5 with {\arrow[#1]{latex}}}}} ,
        worlddot/.style={dotted, line width=.8pt},
	worlddot2/.style={dotted, line width=1pt}   }

\DeclareFontFamily{OT1}{pzc}{}
\DeclareFontShape{OT1}{pzc}{m}{it}{<-> s * [1.350] pzcmi7t}{}
\DeclareMathAlphabet{\mathpzc}{OT1}{pzc}{m}{it}

\def\eps{\epsilon}

\def\d{\mathrm{d}}

\def\eE{\mathrm{E}}
\def\eK{\mathrm{K}}

\renewcommand{\i}{\ensuremath{\mathrm{i}}}

\def\dd{\delta\!\!\!{}^-\!}

\def\d{\mathrm{d}}
\def\eps{\epsilon}

\def\nn{\nonumber}

\def\eqn#1{Eq.~\eqref{#1}}

\def\Eqn#1{Eq.~\eqref{#1}}

\def\Rcite#1{Ref.~\cite{#1}}

\newcommand*\Bell{\ensuremath{\boldsymbol\ell}}

\newcommand{\vev}[1]{\langle #1\rangle}

\newcommand{\widebar}{\overline}

\newcommand{\be}{\begin{equation}}
\newcommand{\ee}{\end{equation}}
\newcommand{\ba}{\begin{align}}
\newcommand{\ea}{\end{align}}
\ifx\genfrac\sdflkaj\else\fi

\newcommand{\mn}{{\mu\nu}}

\newcommand{\pin}{p_{\infty}}

\begin{document}
\preprint{HU-EP-24/08-RTG}
\preprint{CERN-TH-2024-032}

\title{Conservative Black Hole Scattering at  Fifth Post-Minkowskian \\ and First Self-Force Order}

\author{Mathias Driesse\,\orcidlink{0000-0002-3983-5852}} 
\affiliation{%
Institut f\"ur Physik und IRIS Adlershof, Humboldt-Universit\"at zu Berlin,
10099 Berlin, Germany
}

\author{Gustav Uhre Jakobsen\,\orcidlink{0000-0001-9743-0442}} 
\affiliation{%
Institut f\"ur Physik und IRIS Adlershof, Humboldt-Universit\"at zu Berlin,
10099 Berlin, Germany
}
 \affiliation{Max Planck Institut f\"ur Gravitationsphysik (Albert Einstein Institut), 14476 Potsdam, Germany}

\author{Gustav Mogull\,\orcidlink{0000-0003-3070-5717}}
\affiliation{%
Institut f\"ur Physik und IRIS Adlershof, Humboldt-Universit\"at zu Berlin,
10099 Berlin, Germany
} \affiliation{Max Planck Institut f\"ur Gravitationsphysik (Albert Einstein Institut), 14476 Potsdam, Germany}
 
 \author{Jan Plefka\,\orcidlink{0000-0003-2883-7825}} 
\affiliation{%
Institut f\"ur Physik und IRIS Adlershof, Humboldt-Universit\"at zu Berlin,
10099 Berlin, Germany
}

\author{Benjamin Sauer\,\orcidlink{0000-0002-2071-257X}} 
\affiliation{%
Institut f\"ur Physik und IRIS Adlershof, Humboldt-Universit\"at zu Berlin,
10099 Berlin, Germany
}

\author{Johann Usovitsch\,\orcidlink{0000-0002-3542-2786}} 
\affiliation{%
Theoretical Physics Department, CERN, 1211 Geneva, Switzerland
}

\begin{abstract}
We compute the 5PM order contributions to the scattering angle and impulse of classical black hole scattering in the conservative sector
at first self-force order (1SF) using the
worldline quantum field theory formalism. This challenging four-loop computation required the use of advanced integration-by-parts
and differential equation technology implemented on high-performance computing systems.
Use of partial fraction identities allowed us to render the complete integrand in a fully planar form. The
resulting function space is simpler than expected: in the scattering angle we see only
multiple polylogarithms up to weight three, and a total absence of the elliptic integrals
that appeared at 4PM order.
All checks on our result, both 
internal --- cancellation of dimensional regularization poles, preservation of the on-shell condition ---
and external --- matching the slow-velocity limit with the 
post-Newtonian (PN) literature up to 5PN order and matching the tail terms to the 4PM
loss of energy --- are passed.
\end{abstract}
 
\maketitle 

Binary black hole (BH) and neutron star (NS) mergers are today routinely observed by the LIGO-Virgo-KAGRA gravitational wave detectors \cite{LIGOScientific:2016aoc,LIGOScientific:2017vwq,LIGOScientific:2021djp}. With the advent of the third generation of gravitational wave detectors \cite{LISA:2017pwj,Punturo:2010zz,Ballmer:2022uxx},
and LISA's recent approval by the European Space Agency,
we anticipate an experimental accuracy 
increase that will enable unprecedented insights into gravitational, astrophysical, nuclear, and fundamental physics. 
From these experimental programs emerges the theoretical imperative to reach utmost precision in  the gravitational waveforms emitted by these violent cosmic events. To meet this demand, a synergy of perturbative analytical and numerical approaches
is needed to solve the classical general relativistic two-body problem.
The former encompasses the post-Newtonian (PN)~\cite{Blanchet:2013haa,Porto:2016pyg,Levi:2018nxp}
(weak gravitational fields and non-relativistic velocities) and post-Minkowskian (PM)~\cite{Kosower:2022yvp,Bjerrum-Bohr:2022blt,Buonanno:2022pgc,DiVecchia:2023frv,Jakobsen:2023oow} (weak fields) expansions;
the latter encompasses modern numerical relativity~\cite{Pretorius:2005gq,Boyle:2019kee,Damour:2014afa}.
Gravitational self-force (SF) (small mass ratio)~\cite{Mino:1996nk,Poisson:2011nh,Barack:2018yvs,Gralla:2021qaf},
meanwhile, is a hybrid approach: the perturbative SF equations typically being solved numerically.
On the analytical side, the incorporation of perturbative quantum field theory (QFT) techniques
has significantly strengthened this program, most recently within the PM expansion.

In the PM regime, which aligns closely with considerations in particle physics,
the focus is shifted from the merger to the gravitational scattering of two BHs or NSs  \cite{Kovacs:1978eu,Westpfahl:1979gu,Bel:1981be,Damour:2017zjx,Hopper:2022rwo}. The compact bodies are modeled 
as massive point particles interacting through gravity --- an effective worldline description motivated by the scale separation between the intrinsic sizes of the objects ($\sim Gm$) and their separation ($\sim|b|$) \cite{Goldberger:2004jt}. Leveraging this effective worldline approach, key observables in classical two-body scattering --- including the impulse (change of momentum), scattering angle and
far-field waveform --- have been systematically computed to 
high orders in the PM expansion, organized in powers of Newton's constant $G$~\cite{Kalin:2020mvi,Kalin:2020fhe,Kalin:2020lmz,Mogull:2020sak,Jakobsen:2021smu,Dlapa:2021npj,Dlapa:2021vgp,Mougiakakos:2021ckm,Riva:2021vnj,Dlapa:2022lmu,Dlapa:2023hsl}.
Spin and tidal effects have also been incorporated~\cite{Liu:2021zxr,Mougiakakos:2022sic,Riva:2022fru,Jakobsen:2021lvp,Jakobsen:2021zvh,Jakobsen:2022fcj,Jakobsen:2022zsx,Jakobsen:2022psy,Shi:2021qsb,Bastianelli:2021nbs,Comberiati:2022cpm,Wang:2022ntx,Ben-Shahar:2023djm,Bhattacharyya:2024aeq,Jakobsen:2023ndj,Jakobsen:2023hig,Jakobsen:2023pvx}. Complementary perturbative QFT strategies, rooted in scattering amplitudes, have also received considerable attention and
achieved comparable precision~\cite{Neill:2013wsa,Luna:2017dtq,Kosower:2018adc,Cristofoli:2021vyo,Bjerrum-Bohr:2013bxa,Bjerrum-Bohr:2018xdl,Bern:2019nnu,Bern:2019crd,Bjerrum-Bohr:2021wwt,Cheung:2020gyp,Bjerrum-Bohr:2021din,DiVecchia:2020ymx,DiVecchia:2021bdo,DiVecchia:2021ndb,DiVecchia:2022piu,Heissenberg:2022tsn,Damour:2020tta,Herrmann:2021tct,Damgaard:2019lfh,Damgaard:2019lfh,Damgaard:2021ipf,Damgaard:2023vnx,Aoude:2020onz,AccettulliHuber:2020dal,Brandhuber:2021eyq,Bern:2021dqo,Bern:2021yeh,Bern:2022kto,Bern:2023ity,Damgaard:2023ttc,Brandhuber:2023hhy,Brandhuber:2023hhy,Brandhuber:2023hhl,DeAngelis:2023lvf,Herderschee:2023fxh,Caron-Huot:2023vxl,FebresCordero:2022jts,Bohnenblust:2023qmy} ---
see in particular \Rcite{Bern:2023ccb} for related work in electrodynamics.

The present state-of-the-art is 4PM ($G^{4}$),
i.e.~next-to-next-to-next-to-leading order (N$^3$LO),
for the scattering angle and impulse \cite{Bern:2021dqo,Bern:2021yeh,Dlapa:2022lmu,Dlapa:2023hsl,Jakobsen:2023ndj,Jakobsen:2023hig,Jakobsen:2023pvx,Damgaard:2023ttc}.
Determination of these observables required the computation of three-loop, one-parameter Feynman integrals.
Including spin degrees of freedom --- parametrized by the ring radius 
$a=S/m$  --- yields a double expansion for the
impulse and spin-kick as $G^{n_{1}}a^{n_{2}}$. 
Here we have knowledge of the terms up to $(n_{1},n_{2}) = \{(1,\infty)$\cite{Vines:2017hyw}$,(2,5)$\cite{Bern:2022kto,Aoude:2023vdk}$,(3,2)$\cite{Jakobsen:2022fcj,Jakobsen:2022zsx}$,(4,1)$\cite{Jakobsen:2023ndj,Jakobsen:2023hig}$ \}$.
As Kerr-BHs obey the inequality $a\leq Gm$, the physical PM counting in (effective) powers of $G$ adds
$n_{1}+n_{2}$. Hence we presently have the complete knowledge of the scattering observables for Kerr-BHs up to and including the physical 4PM
order. In order to advance to 5PM order, we lack only $(5,0)$, i.e.~the spin-free four-loop contribution.
\begin{figure*}[ht!]
  \begin{subfigure}{0.16\textwidth}
  \centering
  \begin{tikzpicture}[baseline={([yshift=-1ex]current bounding box.south)},scale=.7]
   \coordinate (inA) at (0.3,.7);
    \coordinate (outA) at (3.6,.7);
    \coordinate (inB) at (0.3,-.7);
    \coordinate (outB) at (3.6,-.7);
    \coordinate (xA) at (0.6,.7);
    \coordinate (yA) at (1.5,.7);
    \coordinate (zA) at (2.4,.7);
    \coordinate (wA) at (3.3,.7);
    \coordinate (xB) at (0.6,-.7);
    \coordinate (yB) at (1.5,-.7);
    \coordinate (zB) at (2.4,-.7);
    \coordinate (wB) at (3.3,-.7);
    \coordinate (xM) at (0.6,0);
    \coordinate (yM) at (1.5,0);
    \coordinate (zM) at (2.4,0);
    \coordinate (wM) at (3.3,0);
    \coordinate (xyMB) at (1.05,-.35);
    \coordinate (yzMB) at (1.95,-.35);
    \coordinate (zwMB) at (2.85,-.35);
    \coordinate (yzMA) at (1.95,.35);
    \draw [dotted] (inA) -- (outA);
    \draw [dotted] (inB) -- (outB);
    \draw [photon] (xA) -- (xB) node [midway,left] {$q\,\,\,\,\,$} node [midway,left] {\mbox{\large$\uparrow$}};
    \draw [photon] (wA) -- (wB);
    \draw [photon,red] (xM) -- (yM);
    \draw [photon,red] (yM) -- (zM);
    \draw [photon,red] (zM) -- (wM);
    \draw [photon] (yB) -- (yM);
    \draw [photon] (zB) -- (zM);
    \draw [fill] (xA) circle (.08);
    \draw [fill] (wA) circle (.08);
    \draw [fill] (xB) circle (.08);
    \draw [fill] (yB) circle (.08);
    \draw [fill] (zB) circle (.08);
    \draw [fill] (wB) circle (.08);  
    \draw [fill] (xM) circle (.08);
    \draw [fill] (yM) circle (.08);
    \draw [fill] (zM) circle (.08);  
    \draw [fill] (wM) circle (.08);
    \node at (xyMB) {\mbox{\Large$\circlearrowright$}};
    \node at (xyMB) {\mbox{\scriptsize$\ell_2$}};
    \node at (yzMB) {\mbox{\Large$\circlearrowright$}};
    \node at (yzMB) {\mbox{\scriptsize$\ell_3$}};
    \node at (zwMB) {\mbox{\Large$\circlearrowright$}};
    \node at (zwMB) {\mbox{\scriptsize$\ell_4$}};
    \node at (yzMA) {\mbox{\Large$\circlearrowright$}};
    \node at (yzMA) {\mbox{\scriptsize$\ell_1$}};
  \end{tikzpicture}
  \caption{}
  \end{subfigure}
  \begin{subfigure}{0.16\textwidth}
  \centering
  \begin{tikzpicture}[baseline={([yshift=-1ex]current bounding box.south)},scale=.7]
   \coordinate (inA) at (0.3,.7);
    \coordinate (outA) at (3.6,.7);
    \coordinate (inB) at (0.3,-.7);
    \coordinate (outB) at (3.6,-.7);
    \coordinate (1M) at (0.6,0);
    \coordinate (2M) at (1.3,0);
    \coordinate (3M) at (2.0,0);
    \coordinate (4M) at (2.7,0);
    \coordinate (5M) at (3.4,0);
    \coordinate (xA) at (0.6,.7);
    \coordinate (yA) at (1.5,.7);
    \coordinate (zA) at (2.4,.7);
    \coordinate (wA) at (3.3,.7);
    \coordinate (xB) at (0.6,-.7);
    \coordinate (yB) at (1.5,-.7);
    \coordinate (zB) at (2.4,-.7);
    \coordinate (wB) at (3.3,-.7);
    \coordinate (xM) at (0.6,0);
    \coordinate (yM) at (1.5,0);
    \coordinate (zM) at (2.4,0);
    \coordinate (wM) at (3.3,0);
    \coordinate (xyMB) at (1.05,-.35);
    \coordinate (yzMB) at (1.95,-.35);
    \coordinate (zwM) at (2.85,0);
    \coordinate (yMA) at (1.5,.35);
    \draw [dotted] (inA) -- (outA);
    \draw [dotted] (inB) -- (outB);
    \draw [photon] (xA) -- (xB);
    \draw [photon] (wA) -- (wB);
    \draw [photon,red] (xM) -- (zM);
    \draw [photon] (yB) -- (yM);
    \draw [photon] (zB) -- (zM);
    \draw [photon,red] (zM) -- (zA);
    \draw [zUndirected] (zA) -- (wA);
    \draw [fill] (xA) circle (.08);
    \draw [fill] (zA) circle (.08);
    \draw [fill] (wA) circle (.08);
    \draw [fill] (xB) circle (.08);
    \draw [fill] (yB) circle (.08);
    \draw [fill] (zB) circle (.08);
    \draw [fill] (wB) circle (.08);  
    \draw [fill] (xM) circle (.08);
    \draw [fill] (yM) circle (.08);
    \draw [fill] (zM) circle (.08);  
    \node at (xyMB) {\mbox{\Large$\circlearrowright$}};
    \node at (xyMB) {\mbox{\scriptsize$\ell_2$}};
    \node at (yzMB) {\mbox{\Large$\circlearrowright$}};
    \node at (yzMB) {\mbox{\scriptsize$\ell_3$}};
    \node at (zwM) {\mbox{\Large$\circlearrowright$}};
    \node at (zwM) {\mbox{\scriptsize$\ell_4$}};
    \node at (yMA) {\mbox{\Large$\circlearrowright$}};
    \node at (yMA) {\mbox{\scriptsize$\ell_1$}};
  \end{tikzpicture}
  \caption{}
  \end{subfigure}
  \begin{subfigure}{0.15\textwidth}
  \centering
  \begin{tikzpicture}[baseline={([yshift=-1ex]current bounding box.south)},scale=.7]
   \coordinate (inA) at (0.3,.7);
    \coordinate (outA) at (3.6,.7);
    \coordinate (inB) at (0.3,-.7);
    \coordinate (outB) at (3.6,-.7);
    \coordinate (xA) at (0.6,.7);
    \coordinate (yA) at (1.5,.7);
    \coordinate (zA) at (2.4,.7);
    \coordinate (wA) at (3.3,.7);
    \coordinate (xB) at (0.6,-.7);
    \coordinate (yB) at (1.5,-.7);
    \coordinate (zB) at (2.4,-.7);
    \coordinate (wB) at (3.3,-.7);
    \coordinate (xM) at (0.6,0);
    \coordinate (yM) at (1.5,0);
    \coordinate (zM) at (2.4,0);
    \coordinate (wM) at (3.3,0);
    \coordinate (xyM) at (1.05,0);
    \coordinate (zwM) at (2.85,0);
    \coordinate (yzMA) at (1.95,.35);
    \coordinate (yzMB) at (1.95,-.35);
    \draw [dotted] (inA) -- (outA);
    \draw [dotted] (inB) -- (outB);
    \draw [photon] (xA) -- (xB);
    \draw [photon] (wA) -- (wB);
    \draw [photon,red] (yM) -- (zM);
    \draw [photon,red] (yM) -- (yA);
    \draw [photon] (yB) -- (yM);
    \draw [photon] (zB) -- (zM);
    \draw [photon,red] (zM) -- (zA);
    \draw [zUndirected] (zA) -- (wA);
    \draw [zUndirected] (xA) -- (yA);
    \draw [fill] (xA) circle (.08);
    \draw [fill] (yA) circle (.08);
    \draw [fill] (zA) circle (.08);
    \draw [fill] (wA) circle (.08);
    \draw [fill] (xB) circle (.08);
    \draw [fill] (yB) circle (.08);
    \draw [fill] (zB) circle (.08);
    \draw [fill] (wB) circle (.08);  
    \draw [fill] (yM) circle (.08);
    \draw [fill] (zM) circle (.08); 
    \node at (xyM) {\mbox{\Large$\circlearrowright$}};
    \node at (xyM) {\mbox{\scriptsize$\ell_2$}};
    \node at (zwM) {\mbox{\Large$\circlearrowright$}};
    \node at (zwM) {\mbox{\scriptsize$\ell_4$}};
    \node at (yzMA) {\mbox{\Large$\circlearrowright$}};
    \node at (yzMA) {\mbox{\scriptsize$\ell_1$}};
    \node at (yzMB) {\mbox{\Large$\circlearrowright$}};
    \node at (yzMB) {\mbox{\scriptsize$\ell_3$}}; 
  \end{tikzpicture}
  \caption{}
  \end{subfigure}
  \begin{subfigure}{0.15\textwidth}
  \centering
  \begin{tikzpicture}[baseline={([yshift=-1ex]current bounding box.south)},scale=.7]
   \coordinate (inA) at (0.3,.7);
    \coordinate (outA) at (3.6,.7);
    \coordinate (inB) at (0.3,-.7);
    \coordinate (outB) at (3.6,-.7);
    \coordinate (xA) at (0.6,.7);
    \coordinate (yA) at (1.5,.7);
    \coordinate (zA) at (2.4,.7);
    \coordinate (wA) at (3.3,.7);
    \coordinate (xB) at (0.6,-.7);
    \coordinate (yB) at (1.5,-.7);
    \coordinate (zB) at (2.4,-.7);
    \coordinate (wB) at (3.3,-.7);
    \coordinate (xM) at (0.6,0);
    \coordinate (yM) at (1.5,0);
    \coordinate (zM) at (2.4,0);
    \coordinate (wM) at (3.3,0);
    \coordinate (xyMB) at (1.05,-.35);
    \coordinate (xyMA) at (1.05,.35);
    \coordinate (yzM) at (1.95,0);
    \coordinate (zwM) at (2.85,0);
    \draw [dotted] (inA) -- (outA);
    \draw [dotted] (inB) -- (outB);
    \draw [photon] (xA) -- (xM);
    \draw [photon] (xM) -- (xB);
    \draw [photon,red] (yA) -- (yM);
    \draw [photon] (yM) -- (yB);
    \draw [photon] (zA) -- (zB);
    \draw [photon] (wA) -- (wB);
    \draw [photon,red] (xM) -- (yM);
    \draw [zUndirected] (yA) -- (wA);
    \draw [fill] (xA) circle (.08);
    \draw [fill] (yA) circle (.08);
    \draw [fill] (zA) circle (.08);
    \draw [fill] (wA) circle (.08);
    \draw [fill] (xB) circle (.08);
    \draw [fill] (yB) circle (.08);
    \draw [fill] (zB) circle (.08);
    \draw [fill] (wB) circle (.08);  
    \draw [fill] (xM) circle (.08);
    \draw [fill] (yM) circle (.08);
    \node at (xyMB) {\mbox{\Large$\circlearrowright$}};
    \node at (xyMB) {\mbox{\scriptsize$\ell_2$}};
    \node at (xyMA) {\mbox{\Large$\circlearrowright$}};
    \node at (xyMA) {\mbox{\scriptsize$\ell_1$}};
    \node at (yzM) {\mbox{\Large$\circlearrowright$}};
    \node at (yzM) {\mbox{\scriptsize$\ell_3$}};
    \node at (zwM) {\mbox{\Large$\circlearrowright$}};
    \node at (zwM) {\mbox{\scriptsize$\ell_4$}};
  \end{tikzpicture}
  \caption{}
  \end{subfigure}
  \begin{subfigure}{0.17\textwidth}
  \centering
  \begin{tikzpicture}[baseline={([yshift=-1ex]current bounding box.south)},scale=.7]
   \coordinate (inA) at (0.3,.7);
    \coordinate (outA) at (4.5,.7);
    \coordinate (inB) at (0.3,-.7);
    \coordinate (outB) at (4.5,-.7);
    \coordinate (1A) at (0.6,.7);
    \coordinate (2A) at (1.5,.7);
    \coordinate (3A) at (2.4,.7);
    \coordinate (4A) at (3.3,.7);
    \coordinate (5A) at (4.2,.7);
    \coordinate (1B) at (0.6,-.7);
    \coordinate (2B) at (1.5,-.7);
    \coordinate (3B) at (2.4,-.7);
    \coordinate (4B) at (3.3,-.7);
    \coordinate (5B) at (4.2,-.7);
    \coordinate (1M) at (0.6,0);
    \coordinate (2M) at (1.5,0);
    \coordinate (3M) at (2.4,0);
    \coordinate (4M) at (3.3,0);
    \coordinate (5M) at (4.2,0);
    \coordinate (12M) at (1.05,0);
    \coordinate (23M) at (1.95,0);
    \coordinate (34M) at (2.85,0);
    \coordinate (45M) at (3.75,0);
    \draw [dotted] (inA) -- (outA);
    \draw [dotted] (inB) -- (outB);
    \draw [photon] (1A) -- (1B);
    \draw [photon,red] (2A) -- (2B);
    \draw [photon,red] (3A) -- (3B);
    \draw [photon] (4A) -- (4B);
    \draw [photon] (5A) -- (5B);
    \draw [zUndirected] (1A) -- (2A);
    \draw [zUndirected] (3A) -- (5A);
    \draw [zUndirected] (2B) -- (3B);
    \draw [fill] (1A) circle (.08);
    \draw [fill] (2A) circle (.08);
    \draw [fill] (3A) circle (.08);
    \draw [fill] (4A) circle (.08);
    \draw [fill] (5A) circle (.08);
    \draw [fill] (1B) circle (.08);
    \draw [fill] (2B) circle (.08);
    \draw [fill] (3B) circle (.08);
    \draw [fill] (4B) circle (.08);  
    \draw [fill] (5B) circle (.08);
    \node at (12M) {\mbox{\Large$\circlearrowright$}};
    \node at (12M) {\mbox{\scriptsize$\ell_2$}};
    \node at (23M) {\mbox{\Large$\circlearrowright$}};
    \node at (23M) {\mbox{\scriptsize$\ell_1$}};
    \node at (34M) {\mbox{\Large$\circlearrowright$}};
    \node at (34M) {\mbox{\scriptsize$\ell_3$}};
    \node at (45M) {\mbox{\Large$\circlearrowright$}};
    \node at (45M) {\mbox{\scriptsize$\ell_4$}};
  \end{tikzpicture}
  \caption{}\label{figE}
  \end{subfigure}
  \begin{subfigure}{0.17\textwidth}
  \centering
  \begin{tikzpicture}[baseline={([yshift=-1ex]current bounding box.south)},scale=.7]
   \coordinate (inA) at (0.3,.7);
    \coordinate (outA) at (4.5,.7);
    \coordinate (inB) at (0.3,-.7);
    \coordinate (outB) at (4.5,-.7);
    \coordinate (1A) at (0.6,.7);
    \coordinate (2A) at (1.5,.7);
    \coordinate (3A) at (2.4,.7);
    \coordinate (4A) at (3.3,.7);
    \coordinate (5A) at (4.2,.7);
    \coordinate (1B) at (0.6,-.7);
    \coordinate (2B) at (1.5,-.7);
    \coordinate (3B) at (2.4,-.7);
    \coordinate (4B) at (3.3,-.7);
    \coordinate (5B) at (4.2,-.7);
    \coordinate (1M) at (0.6,0);
    \coordinate (2M) at (1.3,0);
    \coordinate (3M) at (2.0,0);
    \coordinate (4M) at (2.7,0);
    \coordinate (5M) at (3.4,0);
    \coordinate (12M) at (1.05,0);
    \coordinate (23M) at (1.95,0);
    \coordinate (34M) at (2.85,0);
    \coordinate (45M) at (3.75,0);
    \draw [dotted] (inA) -- (outA);
    \draw [dotted] (inB) -- (outB);
    \draw [photon] (1A) -- (1B);
    \draw [photon] (2A) -- (2B);
    \draw [photon] (3A) -- (3B);
    \draw [photon,red] (4A) -- (4B);
    \draw [photon] (5A) -- (5B);
    \draw [zUndirected] (1A) -- (4A);
    \draw [zUndirected] (4B) -- (5B);
    \draw [fill] (1A) circle (.08);
    \draw [fill] (2A) circle (.08);
    \draw [fill] (3A) circle (.08);
    \draw [fill] (4A) circle (.08);
    \draw [fill] (5A) circle (.08);
    \draw [fill] (1B) circle (.08);
    \draw [fill] (2B) circle (.08);
    \draw [fill] (3B) circle (.08);
    \draw [fill] (4B) circle (.08);  
    \draw [fill] (5B) circle (.08);
     \node at (12M) {\mbox{\Large$\circlearrowright$}};
     \node at (12M) {\mbox{\scriptsize$\ell_2$}};
     \node at (23M) {\mbox{\Large$\circlearrowright$}};
     \node at (23M) {\mbox{\scriptsize$\ell_3$}};
     \node at (34M) {\mbox{\Large$\circlearrowright$}};
     \node at (34M) {\mbox{\scriptsize$\ell_4$}};
     \node at (45M) {\mbox{\Large$\circlearrowright$}};
     \node at (45M) {\mbox{\scriptsize$\ell_1$}};
  \end{tikzpicture}
  \caption{}\label{figF}
  \end{subfigure}
     \caption{\small
      The six top-level sectors of the four-loop planar integral family \eqref{eq:planarFamily},
      yielding the $m_{1}^{2} m_{2}^{4}$ 5PM-1SF contributions.
      The $\dd(\ell_i\cdot u_i)$ can here be interpreted as cut propagators --- dotted lines,
      which in the WQFT context alternatively denote the background worldlines.
      In this planar four-loop family we have 13 active propagators~\eqref{effprops},
      the active graviton propagators that may become radiative \eqref{PRregions} being depicted in red.}
    \label{fig:TL}
  \end{figure*}
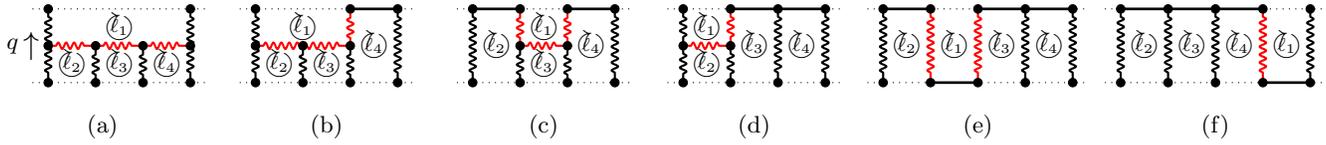

The SF expansion \cite{Mino:1996nk,Poisson:2011nh,Barack:2018yvs,Gralla:2021qaf} is a complementary perturbative scheme in which one assumes a hierarchy in the two BH or NS masses,
$m_{1}\ll m_{2}$, but works \emph{exactly} in $G$. The self-force expansion therefore extends systematically
beyond the geodesic motion of a probe mass moving in the background of a heavy BH or NS. 
One may overlay the PM loop expansion with the SF expansion:
the PM problem factorizes into separate gauge-invariant
SF sectors that may be targeted individually. Concretely, for the 5PM four-loop problem one finds 0SF (known), 1SF (computed here) and 2SF contributions. The complexity of the 
Feynman integrals to be performed grows considerably with the SF order. Moreover, overlaying the PM with the SF expansion for the
scattering scenario is also motivated on astrophysical grounds: statistical estimates for inspirals of stars about super-massive ($M> 10^{6}M_{\odot}$) or intermediate mass  ($M\sim 10^{3}M_{\odot}$) BHs display highly eccentric orbits, potentially observable with
LISA \cite{Amaro-Seoane:2012lgq,Hopman:2005vr,McCart:2021upc}, which may be well captured by PM-improved effective-one-body (EOB) models \cite{Khalil:2022ylj,Rettegno:2023ghr,Buonanno:2024vkx}. 

In this Letter, we compute the previously unknown 5PM contribution at first order in self-force.
Our computation lies at the frontier of present Feynman integration technology.
In order to master it, we optimized on all aspects of this high-precision
challenge: the integrand was produced using the worldline quantum field theory (WQFT)
formalism~\cite{Mogull:2020sak,Jakobsen:2021zvh,Jakobsen:2022psy,Jakobsen:2023oow},
with partial fraction identities used to perform a ``planarization'' prior to integration.
The integration-by-parts (IBP) reduction employed an improved version
of {\tt Kira} \cite{Klappert:2020nbg,Lange:2021edb}.

\sec{Worldline Quantum Field Theory} The non-spinning BHs or NSs are modeled as point particles moving on trajectories $x_{i}^{\mu}(\tau)$. 
In proper time gauge $\dot x_{i}^{2}=1$ the action takes the simple form
\begin{align}\label{eq:action}
S=-\sum_{i=1}^{2}\frac{m_i}2\int\!\d\tau\,g_{\mu\nu}\dot x_i^{\mu}\dot x_i^{\nu}
 - \frac{1}{16\pi G}\int\!\d^{D}x \sqrt{-g} R \, ,
\end{align}
suppressing a gauge-fixing term $S_{\text{gf}}$.
We employ a non-linearly extended de Donder gauge 
that maximally simplifies the three- and four-graviton vertices,
and use dimensional regularization with $D=4-2\epsilon$ in the bulk.
Both the worldline and gravitational fields are expanded about their Minkowskian ($G^{0}$) background configurations
\begin{align}
  \begin{aligned}\label{backgroundexp}
    x_i^\mu &= b_i^\mu \!+\! v_i^\mu \tau \!+\! z_i^\mu\,, \, \quad 
    g_{\mu\nu} = \eta_{\mu\nu} +\sqrt{32\pi G}\, h_\mn \,,
  \end{aligned}
\end{align}
yielding the propagating  worldline deflections $z_i^\mu(\tau)$ and graviton field $h_\mn(x)$.
The incoming data is then spanned by the impact parameter $b^{\mu}=b_{2}^{\mu}-b_{1}^{\mu}$ and
the initial velocities $v^\mu_{1},v^\mu_{2}$, with $v_1^2=v_2^2=1$ and 
$\gamma=v_1\cdot v_2=(1-v^{2})^{-1/2}$.

The quest of solving the equations of motions of \Eqn{eq:action} in a $G$-expansion is solved 
upon quantizing the perturbations $z^{\mu}_{i}$ and $h_{\mu\nu}$:
the tree-level one-point functions then solve the classical equations of motion \cite{Boulware:1968zz}.
The impulse of (say) the first BH, $\Delta p_{1}^{\mu}$, then emerges from 
$\Delta p_{1}^{\mu} = \lim_{\omega\to 0}\omega^{2}\vev{z_{1}^{\mu}(\omega)}$, working in
momentum (energy) space.
The WQFT vertices are given by standard bulk
graviton vertices --- at 5PM we require the $3, 4, 5$ and $6$ graviton vertices --- and worldline vertices 
coupling a single graviton to (0,\ldots, 5)-worldline deflections \cite{Jakobsen:2021zvh,Jakobsen:2023ndj}.
We access the \emph{conservative} sector by employing Feynman propagators  (in-out) in the
bulk and retarded on the worldline (in-in) \cite{Jakobsen:2022psy,Kalin:2022hph}, taking the part real and even in velocity $v$. 
Non-trivial Feynman loop integrals emerge in WQFT due to the hybrid nature of the theory:
the worldlines only conserve the total inflowing energy, as opposed to full four-momentum conservation in the 
bulk. The (non-spinning) $n$th PM contribution to the impulse
is given by $(n-1)$-loop integrals, plus a trivial Fourier transform over the momentum transfer $q$.

\sec{Self-force expansion} The 5PM contribution to the complete impulse, 
$
\Delta p_{1}^{\mu} =\sum_{n=1}^{\infty} G^{n} \Delta p^{(n)\,\mu}
$, factorizes into (effectively) three self-force (SF) contributions:
\begin{align}
\Delta p^{(5) \mu}=  m_{1} m_{2} & \Bigl ( m_2^{4} \Delta p^{(5) \mu}_{\text{0SF}}
+  m_1 m_2^3 \Delta p^{(5) \mu}_{\text{1SF}}   \\
+m_1^2 m_2^2 &\Delta p^{(5) \mu}_{\text{2SF}} 
+ m_1^3 m_2 \Delta p^{(5) \mu}_{\widebar{\text{1SF}}} 
 + m_1^4 \Delta p^{(5) \mu}_{\widebar{\text{0SF}}}\Bigr) \nn \,,
\end{align}
each of which is separately gauge-invariant.
In fact, the SF order may be directly read off a WQFT diagram:
the power of $m_{i}$ is given by the number of times
the $i$'th worldline is ``touched'' ---
see e.g.~Fig.~\ref{fig:TL}, which contains integral graphs belonging to the 1SF ($m_1^2m_2^4$) sector.
Simplest to compute are the probe limit results $\Delta p^{(5) \mu}_{\text{0SF}}$
and $\Delta p^{(5) \mu}_{\widebar{\text{0SF}}}$, which describe geodesic motion
and are known to all orders in $G$ \cite{Damgaard:2022jem}.
They encode the $m_1\ll m_2$ and $m_1\gg m_2$ limits respectively,
and are related to each other by symmetry.
For the conservative dynamics that we focus on here,
the leading (1SF) self-force corrections $\Delta p^{(5) \mu}_{\text{1SF}}$ and
$\Delta p^{(5) \mu}_{\widebar{\text{1SF}}}$ are also related by swapping $1\leftrightarrow 2$.
The conservative 1SF sector result $\Delta p^{(5) \mu}_{\text{1SF}}$
will be a central result of this Letter.

\sec{Integrand generation} The 5PM integrand is generated with  
a Berends-Giele type recursion relation employing the automated vertex rules from the action~\eqref{eq:action},
as discussed in \Rcite{Jakobsen:2023ndj}. It is not a bottleneck of the
computation. 
In the 1SF sector this yields a total of 363 WQFT diagrams;
the probe limit (0SF sector), which we also generate as a test-bed, 
is comprised of 63 diagrams.
After inserting  the Feynman rules using {\tt FORM}~\cite{Ruijl:2017dtg},
$\Delta p^{(5) \mu}$ may be reduced to a sum of scalar-type integrals
by replacing any loop momenta with a free index as \cite{Jakobsen:2022fcj}
\begin{align}
\ell_i^{\mu}\to\sum_{j=1}^{2} (\ell_i\cdot v_{j}) \hat{v}_j^\mu-\frac{ (\ell_i\cdot q)}{|q|^{2}} q^{\mu}\,.
\end{align}
The dual velocities $\hat{v}_1^\mu = (\gamma v_2^{\mu}-v_1^{\mu} )/(\gamma^{2}-1)$ and
$\hat{v}_2^\mu = (\gamma v_1^{\mu}-v_2^{\mu} )/(\gamma^{2}-1)$
satisfy $v_i\cdot\hat{v}_j=\delta_{ij}$.
The momentum impulse is then expressed as linear combinations of scalar integrals
depending trivially on the momentum transfer
$|q|:=\sqrt{-q^2}$ (this being the sole dimensionful quantity in the problem) and non-trivially on $\gamma=v_{1}
\cdot v_{2}$.

In anticipation of the subsequent IBP reduction step,
we organize the resulting scalar integrals into families.
We introduce the following generic 1SF planar integral family,
valid at \emph{any} $L$-loop order:
\begin{subequations}\label{eq:planarFamily}
\begin{align}
  {\cal I}_{\{n\}}^{\{\sigma\}}
  =
  \int_{\ell_1\cdots\ell_L}
  \frac{
    \dd^{(\bar{n}_1\!-\!1)}(\ell_1\cdot v_1)\prod_{i=2}^L\dd^{(\bar{n}_i\!-\!1)}(\ell_i\cdot v_2)
    }{
    \prod_{i=1}^{L}D_{i}^{n_{i}}(\sigma_{i})
    \prod_{I<J}D_{IJ}^{n_{IJ}}
    }\,,
\end{align}
where $\{\sigma\}$ and $\{n\}$ denote collections of $\i 0^+$ signs
and integer powers of propagators respectively.
The worldline propagators $D_i(\sigma_i)$ are %
\begin{align}
  D_1&=\ell_1\cdot v_2+\sigma_1\i 0^+\,, &
  D_{i>1}&=\ell_i\cdot v_1+\sigma_i \i 0^+\,,
\end{align}
and the massless bulk propagators (gravitons) $D_{IJ}$ with $I=(0,i,q)$ are
(suppressing a Feynman $\i 0^{+}$ 
prescription)
\begin{align}\label{DIJs}
  D_{ij}&=(\ell_i-\ell_j)^2
  \,,&
  D_{qi}&=(\ell_i+q)^2
  \,,&
  D_{0i}&=\ell_i^2
  \,.
\end{align}
\end{subequations}
In total we have $L$ linear and $L(L+3)/2$ quadratic propagators at $L$-loop order. 
We also allow for derivatives of the one-dimensional delta function $\dd(\omega):=2\pi\delta(\omega)$:
 \begin{align} \label{eq:delta_derivatives}
   \frac{\dd^{(n)}(\omega)}{(-1)^nn!}=
   \frac{i}{(\omega+\i 0^+)^{n+1}}-\frac{i}{(\omega-\i 0^+)^{n+1}}\,.
 \end{align}
The four-loop family is illustrated in Fig.~\ref{fig:TL},
with the following diagrammatic rules: 
\begin{subequations}
\label{effprops}
  \begin{align}
    \begin{tikzpicture}[baseline={(current bounding box.center)}]
      \begin{feynman}
      \coordinate (x) at (-.9,0);
      \coordinate (y) at (0.3,0);
      \draw [photon] (x) -- (y) node [midway, below] {$k$};
      \draw [fill] (x) circle (.08);
      \draw [fill] (y) circle (.08);
      \end{feynman}
    \end{tikzpicture}\,\,
    \,\,\,&=\frac{1}{k^{2} +\i 0^{+}}\,,\\
    \begin{tikzpicture}[baseline={(current bounding box.center)}]
      \coordinate (in) at (-0.6,0);
      \coordinate (out) at (1.4,0);
      \coordinate (x) at (-.2,0);
      \coordinate (y) at (1.0,0);
      \draw [zParticle] (x) -- (y) node [midway, below] {$k$} node [midway, above] { };
      \draw [background] (in) -- (x);
      \draw [background] (y) -- (out);
      \draw [fill] (x) circle (.08);
      \draw [fill] (y) circle (.08);
    \end{tikzpicture}&=\frac1{k\cdot v_i +\i 0^{+}}\,, \label{zdirprop} \\
    \begin{tikzpicture}[baseline={(current bounding box.center)}]
      \coordinate (in) at (-0.6,0);
      \coordinate (out) at (1.4,0);
      \coordinate (x) at (-.2,0);
      \coordinate (y) at (1.0,0);
      \draw [fill] (x) circle (.08);
      \draw [fill] (y) circle (.08);
      \draw [background] (in) -- (out) node [midway, below] {$k$};
    \end{tikzpicture}&=\dd(k\cdot v_i)\,. \label{dottedprop}
  \end{align}
\end{subequations}
The optional arrow in \eqn{zdirprop} denotes causality flow. 
By interpreting the background worldlines as cut propagators,
we ``close'' the loops of the tree-level WQFT diagrams,
and may thus import the notion of planarity from regular QFT Feynman diagrams.
We note in passing
that this  matches the velocity cuts of Refs.~\cite{Bjerrum-Bohr:2021din,Bjerrum-Bohr:2021wwt}.

\begin{figure}
\begin{tikzpicture}[baseline={([yshift=-1ex]current bounding box.south)},scale=.7]
 \coordinate (inA) at (0.3,.7);
  \coordinate (outA) at (4.5,.7);
  \coordinate (inB) at (0.3,-.7);
  \coordinate (outB) at (4.5,-.7);
  \coordinate (1A) at (0.6,.7);
  \coordinate (2A) at (1.5,.7);
  \coordinate (3A) at (2.4,.7);
  \coordinate (4A) at (3.3,.7);
  \coordinate (5A) at (4.2,.7);
  \coordinate (1B) at (0.6,-.7);
  \coordinate (2B) at (1.5,-.7);
  \coordinate (3B) at (2.4,-.7);
  \coordinate (4B) at (3.3,-.7);
  \coordinate (5B) at (4.2,-.7);
  \coordinate (34M) at (2.85,0);
  \coordinate (1M) at (0.6,0);
  \coordinate (2M) at (1.5,0);
  \coordinate (3M) at (2.4,0);
  \coordinate (4M) at (3.3,0);
  \coordinate (5M) at (4.2,0);
  \coordinate (12M) at (1.05,0);
  \coordinate (23M) at (1.95,0);
  \coordinate (34M) at (2.85,0);
  \coordinate (45M) at (3.75,0);
  \draw [dotted] (inA) -- (outA);
  \draw [dotted] (inB) -- (outB);
  \draw [photon] (1A) -- (1B);
  \draw [photon] (2A) -- (2B);
  \draw [photon] (3A) -- (4B);
  \filldraw[white] (34M) circle (6pt);
  \draw [photon] (4A) -- (3B);
  \draw [photon] (5A) -- (5B);
  \draw [zUndirected] (1A) -- (2A);
  \draw [zUndirected] (3A) -- (5A);
  \draw [zUndirected] (2B) -- (3B);
  \draw [fill] (1A) circle (.08);
  \draw [fill] (2A) circle (.08);
  \draw [fill] (3A) circle (.08);
  \draw [fill] (4A) circle (.08);
  \draw [fill] (5A) circle (.08);
  \draw [fill] (1B) circle (.08);
  \draw [fill] (2B) circle (.08);
  \draw [fill] (3B) circle (.08);
  \draw [fill] (4B) circle (.08);  
  \draw [fill] (5B) circle (.08);
\end{tikzpicture}
\qquad
\begin{tikzpicture}[baseline={([yshift=-1ex]current bounding box.south)},scale=.7]
  \coordinate (inA) at (0.3,.7);
   \coordinate (outA) at (4.5,.7);
   \coordinate (inB) at (0.3,-.7);
   \coordinate (outB) at (4.5,-.7);
   \coordinate (1A) at (0.6,.7);
   \coordinate (2A) at (1.5,.7);
   \coordinate (3A) at (2.4,.7);
   \coordinate (4A) at (3.3,.7);
   \coordinate (5A) at (4.2,.7);
   \coordinate (1B) at (0.6,-.7);
   \coordinate (2B) at (1.5,-.7);
   \coordinate (3B) at (2.4,-.7);
   \coordinate (4B) at (3.3,-.7);
   \coordinate (5B) at (4.2,-.7);
   \coordinate (1M) at (0.6,0);
   \coordinate (2M) at (1.3,0);
   \coordinate (3M) at (2.0,0);
   \coordinate (4M) at (2.7,0);
   \coordinate (5M) at (3.4,0);
   \coordinate (12M) at (1.05,0);
   \coordinate (23M) at (1.95,0);
   \coordinate (34M) at (2.85,0);
   \coordinate (45M) at (3.75,0);
   \draw [dotted] (inA) -- (outA);
   \draw [dotted] (inB) -- (outB);
   \draw [photon] (1A) -- (1B);
   \draw [photon] (2A) -- (2B);
   \draw [photon] (3A) -- (4B);
   \filldraw[white] (34M) circle (6pt);
   \draw [photon] (4A) -- (3B);
   \draw [photon] (5A) -- (5B);
   \draw [zUndirected] (1A) -- (4A);
   \draw [zUndirected] (4B) -- (5B);
   \draw [fill] (1A) circle (.08);
   \draw [fill] (2A) circle (.08);
   \draw [fill] (3A) circle (.08);
   \draw [fill] (4A) circle (.08);
   \draw [fill] (5A) circle (.08);
   \draw [fill] (1B) circle (.08);
   \draw [fill] (2B) circle (.08);
   \draw [fill] (3B) circle (.08);
   \draw [fill] (4B) circle (.08);  
   \draw [fill] (5B) circle (.08);
 \end{tikzpicture}
   \caption{\small
  	Two examples of nonplanar loop integrals.
    By applying the partial-fraction identity~\eqref{eq:PF},
    we may re-express then in terms of the integrals in Figs.~\ref{figE} and \ref{figF} respectively,
    and thus include them in the planar loop integral family~\eqref{eq:planarFamily}.
	   }
  \label{fig:nonplanar}
\end{figure}
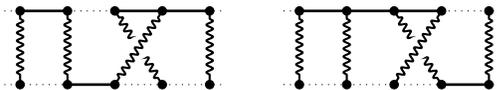

Remarkably, the entire 5PM-1SF result for the momentum impulse
may be expressed in terms of integrals belonging to this planar integral family alone.
To achieve such a representation, graphs with a nonplanar structure ---
such as the two depicted in Fig.~\ref{fig:nonplanar} --- are systematically
eliminated in favor of planar ones.
This is done using partial fraction identities on the worldline propagators:
\begin{align}\label{eq:PF}
  \begin{tikzpicture}[baseline={([yshift=-1ex]current bounding box.center)},scale=.7]
    \coordinate (inA) at (0.2,.6);
    \coordinate (outA) at (3.8,.6);
    \coordinate (inB) at (0.2,-.6);
    \coordinate (outB) at (3.8,-.6);
    \coordinate (xA) at (.8,.6);
    \coordinate (xyA) at (1.5,.6);
    \coordinate (yA) at (2,.6);
    \coordinate (yzA) at (2.5,.6);
    \coordinate (zA) at (3.2,.6);
    \coordinate (xB) at (.8,-.6);
    \coordinate (yB) at (2,-.6);
    \coordinate (zB) at (3.2,-.6);
    \draw [fill] (xA) circle (.08);
    \draw [fill] (yA) circle (.08);
    \draw [fill] (zA) circle (.08);
    \draw [dotted] (inA) -- (outA);
    \draw [zParticle] (xA) -- (yA);
    \draw [zParticle] (zA) -- (yA);
    \draw [photon] (xA) -- (xB);
    \draw [photon] (yA) -- (yB);
    \draw [photon] (zA) -- (zB);
    \draw (xyA) node [above] {$\ell_1\!\cdot\!v_1\,\,\,$};
    \draw (yzA) node [above] {$\,\,\,\ell_2\!\cdot\!v_1$};
    \draw [draw=none] (xB) to[out=-40,in=-140] (zB);
  \end{tikzpicture}\!&=\!
\begin{tikzpicture}[baseline={([yshift=-1ex]current bounding box.center)},scale=.7]
    \coordinate (inA) at (0.2,.6);
    \coordinate (outA) at (3.8,.6);
    \coordinate (inB) at (0.2,-.6);
    \coordinate (outB) at (3.8,-.6);
    \coordinate (xA) at (.8,.6);
    \coordinate (xyA) at (1.5,.6);
    \coordinate (yA) at (2,.6);
    \coordinate (yzA) at (2.5,.6);
    \coordinate (zA) at (3.2,.6);
    \coordinate (xB) at (.8,-.6);
    \coordinate (yB) at (2,-.6);
    \coordinate (zB) at (3.2,-.6);
    \coordinate (yzAB) at (2.6,0);
    \draw [fill] (xA) circle (.08);
    \draw [fill] (yA) circle (.08);
    \draw [fill] (zA) circle (.08);
    \draw [dotted] (inA) -- (outA);
    \draw [zParticle] (xA) -- (yA);
    \draw [zParticle] (yA) -- (zA);
    \draw [photon] (xA) -- (xB);
    \draw [photon] (yA) -- (zB);
    \filldraw[white] (yzAB) circle (6pt);
    \draw [photon] (zA) -- (yB);
    \draw (xyA) node [above] {$\ell_1\!\cdot\!v_1\,\,\,$};
    \draw (yzA) node [above] {$\,\,\,\,\,\ell_{12}\!\cdot\!v_1$};
    \draw [draw=none] (xB) to[out=-40,in=-140] (zB);
  \end{tikzpicture}\,+\,
\begin{tikzpicture}[baseline={([yshift=-1ex]current bounding box.center)},scale=.7]
    \coordinate (inA) at (0.2,.6);
    \coordinate (outA) at (3.8,.6);
    \coordinate (inB) at (0.2,-.6);
    \coordinate (outB) at (3.8,-.6);
    \coordinate (xA) at (.8,.6);
    \coordinate (xyA) at (1.5,.6);
    \coordinate (yA) at (2,.6);
    \coordinate (yzA) at (2.5,.6);
    \coordinate (zA) at (3.2,.6);
    \coordinate (xB) at (.8,-.6);
    \coordinate (yB) at (2,-.6);
    \coordinate (zB) at (3.2,-.6);
    \coordinate (xyAB) at (1.4,0);
    \draw [fill] (xA) circle (.08);
    \draw [fill] (yA) circle (.08);
    \draw [fill] (zA) circle (.08);
    \draw [dotted] (inA) -- (outA);
    \draw [zParticle] (yA) -- (xA);
    \draw [zParticle] (zA) -- (yA);
    \draw [photon] (zA) -- (zB);
    \draw [photon] (yA) -- (xB);
    \filldraw[white] (xyAB) circle (6pt);
    \draw [photon] (xA) -- (yB);
    \draw (xyA) node [above] {$\ell_{12}\!\cdot\!v_1\,\,\,\,\,$};
    \draw (yzA) node [above] {$\,\,\,\ell_2\!\cdot\!v_1$};
    \draw [draw=none] (xB) to[out=-40,in=-140] (zB);
  \end{tikzpicture}\nn\\[-1em]
  \frac1{(\ell_1\cdot v_1)(\ell_2\cdot v_1)}&=
  \frac1{(\ell_1\cdot v_1)(\ell_{12}\cdot v_1)}+
  \frac1{(\ell_{12}\cdot v_1)(\ell_2\cdot v_1)}\,,
\end{align}
where $\ell_{12}^\mu=\ell_1^\mu+\ell_2^\mu$, and
each linear propagator carries an implicit $+\i 0^+$ prescription.
This identity, which may be applied internally within a multi-loop integral
containing linearized propagators,
has the effect of ``untangling'' the crossed bulk propagators,
and can be applied repeatedly in order to produce a fully planar integrand.

\sec{IBP reduction}
The planar integral family~\eqref{eq:planarFamily} splits into two branches:
even ($b$-type) and odd ($v$-type) under the operation $v_i^\mu \rightarrow - v_i^\mu$.
These two branches are thus distinguished by the number of worldline propagators:
even ($b$-type) or odd ($v$-type), including also the number of derivatives on the delta functions. They may be IBP reduced separately,
and in the final answer for the impulse they contribute in the directions of $b^\mu$ and $v_i^\mu$ respectively (hence the name).

Crucially, all $\gamma$-dependence in the integrals lies in the linear propagators and delta functions.
The analytic complexity therefore depends highly on the the combination of contractions with $v_1^\mu$ or $v_2^\mu$ in these propagators.
At $m$SF and $n$PM order we have in the delta functions $m$ loop momenta contracted with $v_1^\mu$ and $n-m-1$ loop momenta contracted with $v_2^\mu$. This yields at 0SF a trivial dependence on $\gamma$ of the integral. At 1SF the functions space becomes more complex due to a single loop momentum being contracted with the velocity of the first worldline. At 2SF order we would have two loop momenta contracted with $v_1^\mu$ and $v_2^\mu$ respectively, and see a significant increase in complexity.

At 5PM-1SF we face four-loop integrals with  13 propagators and 9 irreducible scalar products, cp.~Fig.~\ref{fig:TL}, whose reduction to master integrals poses a significant challenge. 
We use {\tt Kira} \cite{Klappert:2020nbg,Lange:2021edb} to perform this integration-by-parts reduction to master integrals (MIs). We encounter up to nine scalar products in the numerator and up to eight powers (7 dots) of $D$'s in the denominators, i.e.~$n_{i/IJ}\in[-9,8]$ in \eqn{eq:planarFamily}.
The IBP reductions utilize {\tt FireFly} \cite{Klappert:2019emp,Klappert:2020aqs}, a library for reconstructing rational functions from finite field samples generated with {\tt Kira}. 

Several new strategies have been implemented to decrease the runtime of numerical evaluations in an IBP reduction.
The first key concept builds upon the modification of the Laporta algorithm \cite{Laporta:2000dsw}.
For every sector with $n$ absent propagators compared to the top-level-sector we generate equations with the total number of allowed scalar products reduced by $n$.
This approach yields a remarkable $10^{(L-1)}$ runtime improvement compared to the current implementation of the Laporta algorithm in {\tt Kira}. The incorporation of this feature is planned for a future release of {\tt Kira} 3.0 \cite{johann}.

We further observe that the IBP vectors used to formulate equations exhibit a useful feature. 
To reduce a large number of scalar products on linear propagators, it is sufficient for the IBP system to close by seeding at most two scalar products on propagators associated with a graviton.
When reducing a high number of dots on linear propagators it is not necessary to seed dots on the graviton propagators. Implementation of this feature results in an additional factor of 10 in main memory management improvement. The complete IBP reductions took around 300k core hours on HPC clusters. Both the IBP reductions and the impulse were also assembled with the aid of {\tt Kira}.

\sec{Differential equations} 
After IBP reduction we find a total of $236+234$ master integrals (MIs), which are solved using the method of differential equations (DE) \cite{Gehrmann:1999as,Henn:2013pwa}. The needed top sectors of MIs are pictured in Fig.~$\ref{fig:TL}$.
 Grouping them in a vector $\underline{I}$ that obeys  $\frac{\d}{\d x} \underline{I} = M(x,\epsilon)\underline{I}$,
we seek a transformation $\underline{J} = T(x,\epsilon) \underline{I}$ such that the
DE factorizes:
\be
\frac{\d}{\d x} \underline{J} = \epsilon A(x)\underline{J}\,,
\ee
where $x=\gamma -\sqrt{\gamma^2-1}$, which is chosen to rationalize the equation.
To simplify this task it is important to first find a basis in which the dependency on $\gamma$ and $\epsilon$ factorizes \cite{Smirnov:2020quc, Usovitsch:2020jrk}. For this it is helpful to allow for derivatives on the 
delta functions. We employ the algorithms {\tt CANONICA} \cite{Meyer:2017joq}, {\tt INITIAL}\cite{Dlapa:2020cwj}
and {\tt FiniteFlow} \cite{Peraro:2019svx}. More details on these transformations were given in \Rcite{Klemm:2024wtd}. The function space of the integrals of the $(\mathcal{I})$ family
\eqref{eq:planarFamily}, which are needed for the conservative calculation, 
is the same as at 4PM order~\cite{Dlapa:2023hsl}, being comprised of iterated integrals of the integration kernels
$\{\frac 1 x , \frac 1 {1\pm x},\frac{x}{ 1+x^{2}}\}$ plus kernels containing $\eK(1-x^{2})^2$,
$\eK$ being the complete elliptic integral of the first kind.

\sec{Boundary integrals} From the solution of the $\epsilon$-factorized DE the master integrals are determined up to integration
constants. We fix these in the static limit ($\gamma\to 1$, $v\to 0$) using the \emph{method of regions} \cite{Beneke:1997zp,Smirnov:2012gma,Becher:2014oda} by expanding
the integrand in $v$. The regions are characterized by different relative scalings of the bulk graviton loop-momenta
of their spacial and timelike components:
\be\label{PRregions}
	  \ell^\text{P}_i=(\ell_i^0,\Bell_i)\sim(v,1)\,,  \qquad
	  \ell_i^\text{R}=(\ell_i^0,\Bell_i)\sim(v,v)\, .
\ee
referred to as potential (P) or radiative (R) modes. Only gravitons which may go on-shell can turn radiative, i.e.~the
three propagators $\{D_{12},D_{13},D_{14}\}$ of \eqn{DIJs}, typeset in red in
Fig.~$\ref{fig:TL}$. We hence encounter
four possible regions (PPP), (PPR), (PRR) and (RRR). Our definition of conservative dynamics
involves taking the even-in-velocity part, hence we consider only the (PPP) and
(PRR) regions which have this scaling. 
The (PRR) region comes with an overall velocity scaling of $(1-x)^{-4 \epsilon}$,
which accounts for the tail effect and all $\log(1-x)$ functions in the final result. The $236+234$ MIs reduce after IBP reduction
of their static limits to only $2+1$ boundary integrals in the (PRR) and 14+12 integrals in the (PPP) region. 
We solve the (PPP) integrals up to cuts by applying \eqn{eq:delta_derivatives} in reverse; partial fraction identities then constrain their values, making them expressible in terms of $\Gamma$ functions. Interestingly, two-worldline integrals are not fully constrained by this approach, yet appear in linear combinations such that the unknown factor cancels out in the final result. We are not able to reduce the (PRR) integrals using cuts and their expressions are more complicated, involving hypergeometric functions.

\sec{Function space} Surprisingly, the resulting function space is simpler than anticipated. The answer for $\Delta p^{(5) \mu}_{\text{1SF}}$ in the $b^\mu$ direction is given by multiple polylogarithms (MPLs)~\cite{goncharov2011multiple,goncharov2001multiple,Duhr:2019tlz}
up to weight 3.
These MPLs are defined by 
\be
\label{MPLdef}
G(a_{1},\ldots,a_{n};y) = \int_{0}^{y} \frac{\d t}{t-a_{1}} G(a_{2},\ldots,a_{n};t)\,,
\ee
with $G(\vec{0}_n,y)={\log^n({y})}/{n!}$ and $a_{i}, y\in \mathbb{C}$.
Even though we encounter the known elliptic integration kernels in the DEs of these integrals,
they only contribute to the answer at $\mathcal{O} (\epsilon)$,
and thus disappear once we take the limit $D\to4$. In the final result complete
elliptic integrals of the first and second kind appear only in the $v$-direction in the combinations
$\eK(1-x^{2})^{2}$, $\eE(1-x^{2})^{2}$ and $\eE(1-x^{2})\eK(1-x^{2})$. In fact, the $v$-direction
component is entirely determined by lower-order PM results upon momentum conservation. 
As we shall see below, the function space of the scattering angle therefore consists only of MPLs.

\sec{Results}
We begin with the 5PM-1SF momentum impulse $\Delta p^{(5) \mu}_{\text{1SF}}$. It may be decomposed as
\begin{align}
  & \Delta p^{(5) \mu}_{\text{cons,1SF}}  = 
  \frac{1}{|b|^{5}}
  \sum_{\rho=\hat b,\hat{v}_{1},\hat{v}_{2} }
  \rho^{\mu}
     \sum_{\alpha }
    F^{(\rho)}_{\alpha}(\gamma)\,
    d_{ \alpha}^{(\rho)}(\gamma) 
\,,
\end{align}
with the basis vectors $\rho^{\mu}=\{b^{\mu}/|b|,\hat v_{1}^{\mu},\hat v_{2}^{\mu}\}$.
The $  d_{ \alpha}^{(\rho)}(\gamma)$ are rational functions (up to integer powers of
$\sqrt{\gamma^{2}-1}$). For the explicit expressions we refer the reader to the 
ancillary file.
The non-trivial $\gamma$ dependence is spanned by the functions
$  F^{(\rho)}_{\alpha}(\gamma)$ that take the surprisingly simple form %
\begin{align}
 F^{\hat{b}}_{\alpha}(\gamma) &= \{f_{1}(\gamma), \ldots , f_{31}(\gamma) \}
\, , \qquad \gamma_\pm=\gamma\pm1 
\nn \\
F^{\hat v_{1}}_{\alpha}(\gamma) &= \left\{ g_{k}(\gamma),
\eK^2\left [ \frac{\gamma_{-}}{\gamma_{+}} \right ],
   \eE^2\left [ \frac{\gamma_{-}}{\gamma_{+}} \right ],
   \eK\left [ \frac{\gamma_{-}}{\gamma_{+}} \right ]
   \eE\left [ \frac{\gamma_{-}}{\gamma_{+}} \right ]\right\}\!, \nn \\
F^{\hat v_{2}}_{\alpha}(\gamma) &= \{1\}\, ,
\end{align} 
where the 31 functions $f_{k}(\gamma)$ are given by MPLs up to weight three, explicitly stated in Table~\ref{table:Bfunctions} in the supplementary material, and 
$g_{k}(\gamma)$ involve MPLs up to weight two known from the 4PM scattering angle \cite{Jakobsen:2023pvx}.
We choose  to present our results in terms of $y=1-x$,
the five-letter alphabet (shifted with respect to the DEs) then being $\{0,1,2,1 \pm\i\}$.
This avoids a proliferation of $\zeta$-values, and renders the small-velocity expansion more natural. Complex arguments always appear in conjugate combinations, such that the imaginary part cancels.
We also present details on the 0SF computation that was done as a test bed in the supplementary
material. All our results are collected in an ancillary file attached to the \texttt{arXiv}
submission of this Letter.

The conservative scattering angle $\theta_{\text{cons}}$ may be extracted from the impulse using
$|\Delta p^\mu_{i,\rm cons}|=2p_\infty\sin(\theta_{\text{cons}}/2)$.
Here $p_\infty=m_1 m_2 \sqrt{\gamma^2-1}/E$, the total (conserved) energy is $E=M\sqrt{1+2\nu(\gamma-1)}$
and the total mass is $M=m_1+m_2$, with $\nu=m_1m_2/M^2$ the symmetric mass ratio.
The scattering angle may then be double expanded as
\begin{align}
  \label{thetadef}
    \theta_{\text{cons}}=\frac{E}{M} \sum_{n\geq1}\sum_{m=0}^{\lfloor\frac{n-1}{2}\rfloor}
    \bigg(
      \frac{GM}{|b|}
      \bigg)^n
      \nu^m
    \theta^{(n,m)}_{\text{cons}}(\gamma)\,,
  \end{align}
where $n$ denotes the PM, $m$ the SF orders and we use the floor function $\lfloor{.}\rfloor$. The central result of our work is the 5PM-1SF
contribution that takes the form
\begin{align}
  \theta^{(5,1)}_{\text{cons}}
  =
  \sum_{k=1}^{31}
  c_k(\gamma) f_k(\gamma)\,,
\end{align}
where $f_k(\gamma)$ are the linear combinations of MPLs up to weight three  and $c_k(\gamma)$ are polynomials in $\gamma$ except for integer powers of $\sqrt{\gamma^2-1}=\gamma v$ and $\gamma$.
Notice here the total absence of elliptic functions.
Both the functions $f_k(\gamma)$ and the coefficients $c_k$ have a definite
parity under $v\to-v$ such that the angle has even parity (up to factors of $\log(v)$).
They are explicitly stated in Tables~\ref{table:Bfunctions} and \ref{table:Cpols} of the supplementary material.

\sec{Checks} As validation of our result for the impulse $\Delta p_1^{\mu}$, 
the following checks were successfully performed: (1) total momentum conservation $p_{1}^{2}= (p_{1}+\Delta p_{1})^{2}$,
(2) reproduction of the geodesic motion (0SF), (3) agreement in the $v\to0$ limit with the scattering 
angle up to 5PN order~\cite{Bini:2021gat,Blumlein:2021txe}:
\begin{align}
  &\theta^{(5,1)}_{\text{cons}}
  =
  \frac{4}{5v^8}
  -
  \frac{137}{5v^6}
  +
  \frac{41\pi^2}{4 v^4}
  -
  \frac{3427}{6v^4}
  +
    \frac{3593\pi^2}{72v^2}
    -
    \frac{2573399}{2160v^2}
    \nn
    \\
    &\!\!
    + \frac{246527 \pi^2}{1440}-\frac{1099195703}{756000}
    -\frac{128}{45}
    \Big[
      \frac{98}{v^2}
      +
      \frac{59}{35}
      \Big]
    \text{log}[2v]+\cdots
  \end{align}
with the velocity $v=\sqrt{\gamma^2-1}/\gamma$. Finally,
(4) the discontinuity of the scattering angle in the complex plane $\gamma\in\mathbb{C}$
is given by the radiated energy at one order lower in the PM expansion~\cite{Cho:2021arx,Dlapa:2021vgp,Jakobsen:2023hig,Jakobsen:2023pvx,Bini:2017wfr}:
\begin{align}\label{eq:discontinuity}
    \frac{\theta_{\rm cons}(-\gamma_{-}\!+\!\i\eps)-\theta_{\rm cons}(-\gamma_{-} \! -\! \i\eps)}{2i\pi}
    =
      G E
      \frac{\partial E_{\rm rad}|_{{\rm odd-in-}v}}{
        \partial L
      }
  \end{align}
with the total angular momentum $L=\pin |b|$.
This operation picks out the coefficient of $\log(\gamma_-)=\log(\gamma-1)$,
with the branch cut naturally extending along the negative real axis.
Given that it is by definition even-in-$v$,
our conservative angle matches the odd-in-$v$ part of the 4PM radiated energy $E_{\rm rad}$
($L$ being odd in $v$).
Upon including dissipative effects in the scattering angle,
we anticipate a match to the full radiated energy.
With our new 5PM-1SF result, we have verified Eq.~\eqref{eq:discontinuity} to the corresponding order with the 4PM-accurate loss of energy on the right-hand-side~\cite{Dlapa:2022lmu,Damgaard:2023ttc,Jakobsen:2023hig}.

\sec{Outlook}
In this Letter we have computed the first complete results for scattering observables
involving non-spinning black holes and neutron stars at 5PM ($G^5$) order --- the 1SF component.
This was an exceptionally challenging calculation requiring advances in IBP technology plus
high-performance computing.
The biggest surprise, given the appearance of elliptic E/K functions at 4PM order,
was the total \emph{absence} of these terms in the 5PM-1SF scattering angle,
which consists only of MPLs up to weight three.
This happens despite these functions appearing in the corresponding DEs.
Having so far focused on the purely conservative sector,
the question now arises whether this pattern persists when dissipative effects are also included.
It will also be fascinating to see whether the Calabi-Yau three-fold,
which appears in the DE of the dissipative effects \cite{Klemm:2024wtd},
contributes to the full answer.

Looking further ahead, our main challenge will be to complete 5PM with the missing 2SF component.
This represents another leap in complexity.
Nevertheless, it is an important task:
with a complete knowledge of the 5PM scattering dynamics
(including spin, which appears at lower loop orders)
our results will full encapsulate the 4PN conservative two-body dynamics.
Our scattering angle is in one-to-one correspondence
with a hyperbolic two-body Hamiltonian:
given recent promising work on mapping unbound to bound orbits in the presence of tails~\cite{Dlapa:2024cje},
there is a prospect of incorporating our results into future-generation gravitational waveform models.
Resummation into the strong-field regime for scattering events using EOB
\cite{Rettegno:2023ghr,Buonanno:2024vkx}
will also likely show further improvements with respect to NR.

\sec{Acknowledgments}
We thank A.~Klemm and C.~Nega for ongoing collaboration,
A.~Patella for advice on HPC and C.~Dlapa,
G.~K\"alin, Z.~Liu and R.~Porto for discussions and 
important comments.
This work was funded by the Deutsche Forschungsgemeinschaft
(DFG, German Research Foundation)
Projektnummer 417533893/GRK2575 ``Rethinking Quantum Field Theory'' 
and  by the European Union through the 
European Research Council under grant ERC Advanced Grant 101097219 (GraWFTy).
Views and opinions expressed are however those of the authors only and do not necessarily reflect those of the European Union or European Research Council Executive Agency. Neither the European Union nor the granting authority can be held responsible for them.
The authors gratefully acknowledge the computing time granted
at NHR@ZIB.

\clearpage

\appendix
\begin{widetext}
\section*{Supplementary Material}

\sec{Probe limit}
In order to reproduce the probe limit using our formalism, we also computed the 63 Feynman
diagrams contributing to the 5PM-0SF part of the momentum impulse. Following the same
steps on integrand generation and tensor reduction we arrive at the
5PM-0SF probe integral family 
\begin{subequations}
  \label{eq:probe}
  \begin{align}
    (PR)^{(\sigma_1,\sigma_2,\sigma_{3},\sigma_4)}_{n_1,n_2,...,n_{18}}
    =
    \int_{\ell_1,\ell_2,\ell_3,\ell_4}\!\!\!\!\!\!\!\!\!\!\!\!\!\!
    \frac{
      \dd(\ell_1\cdot v_{1})
      \dd(\ell_2\cdot v_{1})
      \dd(\ell_3\cdot v_{1})
       \dd(\ell_4\cdot v_{1})
    }{
      D_1^{n_1}
      D_2^{n_2}
      \cdots
      D_{18}^{n_{18}}
    } \, ,
  \end{align}
  with the propagators ($j=1,2,3,4$ and $\ell_{ij\ldots k}:=\ell_{i}+\ell_{j}+\ldots +\ell_{k}$)
 \begin{align}
    &
    D_j
    =
    \ell_{j}\cdot v_{2} +\sigma_{k}i0^+
    \,,\,
    D_{5}
    =
    (\ell_{1234}+q)^2
     \,,\,
     D_{6}
    =
    (\ell_{123}+q)^2
    \,,\,
    D_{7}
    =
    (\ell_{12}+q)^2
     \,,\,
     D_{8}
    =
    (\ell_{1}+q)^2
    \,,\,\,\\
    &D_{9}
    =
    \ell_{12}^2
    \, , \,
    D_{10}
    =
    \ell_{13}^2
    \, , \,
     D_{11}
    =
    \ell_{23}^2
    \, , \,
     D_{12}
    =
    \ell_{14}^2
    \, , \,
     D_{13}
    =
    \ell_{24}^2
    \, , \,
     D_{14}
    =
    \ell_{34}^2
    \, , \,
    D_{14+j}
    =
    \ell_j^2
   \,.\nn
  \end{align}
  \end{subequations}
A difference with respect to the 1SF calculation was that we did \emph{not} apply partial fraction identities:
while we could, in principle, have reduced to a planar integral family similar to the one used at 1SF~\eqref{eq:planarFamily},
we found it simpler to use the integral family $PR$ instead.
Note that these integrals are independent of $\gamma$ and evaluate to pure functions
of the dimensional regulator $\epsilon$. They were IBP reduced to a set of
$8+7$ MIs using {\tt Kira}.
These MIs 
also feature as boundary integrals in the 1SF computation and needed to be evaluated.
The final result for the probe limit momentum impulse is very compact, and takes the form
\begin{align}
  \Delta p^{(5) \mu}_{\text{0SF}}  =  \frac{1}{|b|^{5}}\Bigl (&
\frac{32 \left(192 \gamma ^{10}-672 \gamma ^8+832 \gamma ^6-424 \gamma ^4+74 \gamma ^2-1\right)-9 \pi ^2 \left(2 \gamma ^2-1\right) \left(5
   \gamma ^4-6 \gamma ^2+1\right)^2}{16 \left(\gamma ^2-1\right)^{9/2} }\frac{b^{\mu}}{|b|}\nn\\
   & +
   \frac{3 \pi  \left(3590 \gamma ^8-7541 \gamma ^6+4907 \gamma ^4-1071 \gamma ^2+51\right) }{32
   \left(\gamma ^2-1\right)^3 }
   \,  \hat v_{1}^{\mu}\Bigr )\, .
\end{align}
The resulting scattering angle agrees  with the result of \Rcite{Damgaard:2022jem}
at 5PM-0SF order.

\sec{Basis functions and coefficients}
In Table~\ref{table:Bfunctions} we list the 31 basis functions that span the
5PM-1SF scattering angle  \eqref{MPLdef}. They are polynomials in MPLs
based on the alphabet $\{0,1,2,1\pm\i\}$ and pure weight functions of weight $0,1,2$
and 3  with multiplicities $1,2,6$ and 22 respectively. In addition they possess a definite parity under $v\to -v$ for which
$\sqrt{\gamma^{2}-1}=\gamma v$ changes sign. In Table~\ref{table:Cpols}
we list the corresponding coefficient polynomials
of \eqn{thetadef} for the 5PM-1SF scattering angle. For an efficient numerical evaluation of MPLs see
{\tt PolyLogTools} \cite{Duhr:2019tlz} using {\tt GiNaC} \cite{Bauer:2000cp}.

\begin{table*}[h!]
	\setlength{\tabcolsep}{1pt} 
	\renewcommand{\arraystretch}{3}
	\begin{tabular}{|c|}
		\hline
		\scalebox{0.74}{\tabeq{23cm}{ \\[-0.3cm]
f_{1}(\gamma) =&1\cr
f_{2}(\gamma) =&G(1;y)\cr
f_{3}(\gamma) =&2 (G(0;y)-G(1;y)+G(2;y)+\log (2))\cr
f_{4}(\gamma) =&\pi ^2\cr
f_{5}(\gamma) =&G(1;y)^2\cr
f_{6}(\gamma) =&2 G(1;y) (G(0;y)-G(1;y)+G(2;y)+\log (2))\cr
f_{7}(\gamma) =&\frac{1}{2} G(1;y)^2+(-G(1;y)+G(1-\i;y)+G(1+\i;y)) G(1;y)-G(1,1-\i;y)-G(1,1+\i;y)\cr
f_{8}(\gamma) =&-\frac{1}{2} G(1;y)^2+G(2;y) G(1;y)+G(0,1;y)-G(1,2;y)\cr
f_{9}(\gamma) =&-G(1;y) G(2;y)+G(0,1;y)+G(1,2;y)\cr
f_{10}(\gamma) =&\pi ^2 G(1;y)\cr
f_{11}(\gamma) =&-G(1;y) \left(-G(1;y)^2-2 G(0;y) G(1;y)+G(2;y) G(1;y)-\log (4) G(1;y)-G(0,1;y)+4 G(1,1-\i;y)+4 G(1,1+\i;y)-G(1,2;y)\right)\cr
f_{12}(\gamma) =&-G(1;y)^3+2 G(2;y) G(1;y)^2+(G(1;y) G(2;y)-G(0,1;y)-G(1,2;y)) G(1;y)+4 \left(-\frac{1}{6} G(1;y)^3+G(1,1,1-\i;y)+G(1,1,1+\i;y)\right)\cr
f_{13}(\gamma) =&\frac{1}{4} G(1;y)^3-G(2;y) G(1;y)^2+G(2;y)^2 G(1;y)-G(0,1;y) G(1;y)+G(1,2;y) G(1;y)+2 G(0,0,1;y)+G(0,1,1;y)-G(1,1,2;y)-2 G(1,2,2;y)\cr
f_{14}(\gamma) =&\left(-\frac{1}{2} G(1;y)^2+G(2;y) G(1;y)+G(0,1;y)-G(1,2;y)\right) (2 G(0;y)-G(1;y)+\log (4))\cr
f_{15}(\gamma) =&-\frac{1}{6} G(1;y)^3+G(1,1-\i;y) G(1;y)+G(1,1+\i;y) G(1;y)+(2 G(2;y)-G(1;y)) \left(-\frac{1}{2} G(1;y)^2+G(1,1-\i;y)+G(1,1+\i;y)\right)-2 G(1,1,1-\i;y)
\\ &
-2 G(1,1,1+\i;y)+2 G(1,1,2;y)-2 G(1,1-\i,2;y)-2 G(1,1+\i,2;y)\cr
f_{16}(\gamma) =&(-G(1;y) G(2;y)+G(0,1;y)+G(1,2;y)) (2 G(0;y)-G(1;y)+\log (4))\cr
f_{17}(\gamma) =&2 \left(-\frac{1}{6} G(1;y)^3+G(0,1,1;y)+G(1,1,2;y)\right)-\frac{3}{2} G(1;y) (-G(1;y) G(2;y)+G(0,1;y)+G(1,2;y))\cr
f_{18}(\gamma) =&\frac{1}{6} G(1;y)^3-(-G(1;y)+G(1-\i;y)+G(1+\i;y)) (2 G(2;y)-G(1;y)) G(1;y)-G(1,1-\i;y) G(1;y)-G(1,1+\i;y) G(1;y)
\\ &
-(G(1;y)-G(1-\i;y)-G(1+\i;y)) (-2 G(0;y)+G(1;y)+\log (4)) G(1;y)-4 (-G(1;y)+G(1-\i;y)+G(1+\i;y)) \log (2) G(1;y)
\\&
+2 G(1,1,1-\i;y)+2 G(1,1,1+\i;y)-2 G(1,1,2;y)+2 G(1,1-\i,2;y)+2 G(1,1+\i,2;y)
\\&
+\left(-\frac{1}{2} G(1;y)^2+G(1,1-\i;y)+G(1,1+\i;y)\right) (2 G(0;y)-G(1;y)-\log (4))+4 \left(-\frac{1}{2} G(1;y)^2+G(1,1-\i;y)+G(1,1+\i;y)\right) \log (2)\cr
f_{19}(\gamma) =&G(1;y) \left(-\frac{1}{2} G(1;y)^2+G(2;y) G(1;y)+G(0,1;y)-G(1,2;y)\right)\cr
f_{20}(\gamma) =&(-G(1;y)+G(1-\i;y)+G(1+\i;y)) \left(2 G(0,1;y)-\frac{1}{2} G(1;y)^2\right)+(-G(1;y)+G(1-\i;y)+G(1+\i;y)) \left(2 (G(1;y) G(2;y)-G(1,2;y))-\frac{1}{2} G(1;y)^2\right)
\\&
-2 (-G(0,1,1;y)+G(0,1,1-\i;y)+G(0,1,1+\i;y))+2 (-G(1;y) G(1,2;y)+2 G(1,1,2;y)+G(1,2,1-\i;y)+G(1,2,1+\i;y))\cr
f_{21}(\gamma) =&2 G(1;y) (G(1;y) G(2;y)-G(0,1;y)-G(1,2;y))\cr
f_{22}(\gamma) =&\frac{1}{2} \Bigl(\frac{1}{6} G(1;y)^3-G(1,1-\i;y) G(1;y)-G(1,1+\i;y) G(1;y)+\Bigl(\frac{1}{2} G(1;y)^2-G(1,1-\i;y)-G(1,1+\i;y)+2 (-G(1;y) G(2;y)+G(1,2;y)+G(2,1-\i;y)
\\ &
+G(2,1+\i;y))\Bigr) G(1;y)+2 (-G(0,1,1;y)+G(0,1-\i,1;y)+G(0,1+\i,1;y))+2 G(1,1,1-\i;y)+2 G(1,1,1+\i;y)\Bigr)\cr
f_{23}(\gamma) =&\frac{1}{12} G(1;y)^3-G(2;y) G(1;y)^2+G(2;y)^2 G(1;y)+G(0,1;y) G(1;y)+G(1,2;y) G(1;y)-2 G(0,0,1;y)-G(0,1,1;y)-G(1,1,2;y)-2 G(1,2,2;y)\cr
f_{24}(\gamma) =&\frac{1}{2} G(1;y)^3-6 G(0,1,1;y)+4 G(0,1,2;y)+8 G(0,2,1;y)-4 (G(1;y) G(1,2;y)-2 G(1,1,2;y))-2 G(1,1,2;y)\cr
f_{25}(\gamma) =&-\frac{1}{12} G(1;y)^3+G(0,1,1;y)-2 G(0,1,2;y)+G(1,1,2;y)\cr
f_{26}(\gamma) =&\frac{1}{2} \Bigl(-\frac{1}{6} G(1;y)^3+G(1,1-\i;y) G(1;y)+G(1,1+\i;y) G(1;y)+\Bigl(\frac{1}{2} G(1;y)^2-G(1,1-\i;y)-G(1,1+\i;y)+2 (-G(1;y) G(2;y)+G(1,2;y)
\\ &
+G(2,1-\i;y)+G(2,1+\i;y))\Bigr) G(1;y)-2 (-G(0,1,1;y)+G(0,1-\i,1;y)+G(0,1+\i,1;y))-2 G(1,1,1-\i;y)-2 G(1,1,1+\i;y)\Bigr)\cr
f_{27}(\gamma) =&\frac{1}{6} G(1;y)^3+\frac{1}{2} (-G(1;y)+G(1-\i;y)+G(1+\i;y))^2 G(1;y)-G(1,1-\i;y) G(1;y)-G(1,1+\i;y) G(1;y)
\\ &
-(-G(1;y)+G(1-\i;y)+G(1+\i;y)) \left(-\frac{1}{2} G(1;y)^2+G(1,1-\i;y)+G(1,1+\i;y)\right)+G(1,1,1-\i;y)+G(1,1,1+\i;y)
\\ &
+G(1,1-\i,1-\i;y)+G(1,1-\i,1+\i;y)+G(1,1+\i,1-\i;y)+G(1,1+\i,1+\i;y)\cr
f_{28}(\gamma) =&G(1;y)^3\cr
f_{29}(\gamma) =&G(1;y)^2 \Bigl(-G(1;y)+G(1-\i;y)+G(1+\i;y)\Bigr)\cr
f_{30}(\gamma) =&-\Bigl(G(1;y)-2 G(2;y)\Bigr) \left(2 G(0,1;y)-\frac{1}{2} G(1;y)^2\right)\cr
f_{31}(\gamma) =& \Bigl(G(1;y)-2 G(2;y) \Bigr) \left(\frac{1}{2} G(1;y)^2-2 (G(1;y) G(2;y)-G(1,2;y))\right)\cr
}}
\\
		\hline
	\end{tabular}
    \caption{Basis functions of the 5PM-1SF scattering angle
     $\theta^{(5,1)}
  =
  \sum_{k=1}^{31}
  c_k(\gamma) f_k(\gamma)$ of \eqn{thetadef}. The $G(a_{1},\ldots, a_{n};y)$ are the multiple polylogarithms (MPLs) defined in  \eqref{MPLdef} and 
    $y=1-x=1 -\gamma +\sqrt{\gamma^2-1}$. Note that MPLs with complex arguments always appear in conjugate pairs securing a real result.
    }
	\label{table:Bfunctions}
\end{table*}

\begin{table*}[h!]
	\setlength{\tabcolsep}{1pt} 
	\renewcommand{\arraystretch}{3}
	\begin{tabular}{|c|}
		\hline
		\scalebox{0.74}{\tabeq{23cm}{\\[-0.2cm]
c_{1}(\gamma) =&\frac{1880064 \gamma ^{19}+1880064 \gamma ^{18}+42654086 \gamma ^{17}+20978054 \gamma ^{16}-305752626 \gamma ^{15}-236079666 \gamma ^{14}+597683406 \gamma ^{13}+516398286 \gamma ^{12}-403178675 \gamma ^{11}}{7560 (\gamma^{2} -1)^4 \gamma ^7 (\gamma +1)} \\ &
+\frac{-362536115 \gamma ^{10}+77856912 \gamma ^9+70236432 \gamma ^8+16701489 \gamma ^7+16955505 \gamma ^6-536235 \gamma ^5-536235 \gamma ^4+393120 \gamma ^3+393120 \gamma ^2+10395 \gamma +10395}{7560 (\gamma^{2} -1)^4 \gamma ^7 (\gamma +1)}\cr
c_{2}(\gamma) =&-\frac{651264 \gamma ^{20}-7809042 \gamma ^{18}-23185512 \gamma ^{16}+169295016 \gamma ^{14}-315460542 \gamma ^{12}+277369170 \gamma ^{10}-134264214 \gamma ^8+6510035 \gamma ^6-988015 \gamma ^4+240905 \gamma ^2-18585}{2520 \gamma ^8 \left(\gamma ^2-1\right)^{9/2}}\cr
c_{3}(\gamma) =&\frac{6144 \gamma ^{16}-587336 \gamma ^{14}+4034092 \gamma ^{12}-417302 \gamma ^{10}-5560073 \gamma ^8-142640 \gamma ^6+35710 \gamma ^4-8250 \gamma ^2+1575}{360 (\gamma^{2} -1)^3 \gamma ^7 }\cr
c_{4}(\gamma) =&-\frac{\gamma  \left(32768 \gamma ^8-90112 \gamma ^6+1564672 \gamma ^4-1872978 \gamma ^2-7817455\right)}{336 (\gamma^{2} -1)^2}\cr
c_{5}(\gamma) =&-\frac{491520 \gamma ^{22}-2482176 \gamma ^{20}+10655064 \gamma ^{18}-32742084 \gamma ^{16}+17085516 \gamma ^{14}+61205662 \gamma ^{12}-59068870 \gamma ^{10}-5433687 \gamma ^8+1352120 \gamma ^6-330890 \gamma ^4+72450 \gamma ^2}{840 (\gamma^{2} -1)^5 \gamma ^7}\\&
+
\frac{11025}{840 (\gamma^{2} -1)^5 \gamma ^7 }\cr
c_{6}(\gamma) =&-\frac{24576 \gamma ^{18}+213480 \gamma ^{16}-1029342 \gamma ^{14}-1978290 \gamma ^{12}+3752006 \gamma ^{10}+816595 \gamma ^8-55260 \gamma ^6+13690 \gamma ^4-3100 \gamma ^2+525}{120 \gamma ^8 \left(\gamma ^2-1\right)^{7/2}}\cr
c_{7}(\gamma) =&\frac{198856 \gamma ^{14}-689664 \gamma ^{12}-154716 \gamma ^{10}+666260 \gamma ^8-5091 \gamma ^6-1935 \gamma ^4+155 \gamma ^2-105}{12 \gamma ^8 \left(\gamma ^2-1\right)^{5/2}}\cr
c_{8}(\gamma) =&-\frac{49152 \gamma ^{18}-208896 \gamma ^{16}+1182464 \gamma ^{14}-3741239 \gamma ^{12}+3040161 \gamma ^{10}+1882567 \gamma ^8-2828161 \gamma ^6+49728 \gamma ^4-2268 \gamma ^2+1260}{42 \gamma ^6 \left(\gamma ^2-1\right)^{7/2}}\cr
c_{9}(\gamma) =&-\frac{\gamma  \left(525 \gamma ^8-450 \gamma ^6+17700 \gamma ^4-12598 \gamma ^2-5369\right)}{4 \left(\gamma ^2-1\right)^{7/2}}\cr
c_{10}(\gamma) =&-\frac{81920 \gamma ^6+189180 \gamma ^4-1240416 \gamma ^2-199207}{48 \left(\gamma ^2-1\right)^{5/2}}\cr
c_{11}(\gamma) =&-\frac{128 \gamma  \left(2 \gamma ^2-3\right) \left(8 \gamma ^6-6 \gamma ^4-51 \gamma ^2-8\right)}{(\gamma^{2} -1)^4 }\cr
c_{12}(\gamma) =&-\frac{\gamma  \left(2 \gamma ^2-3\right) \left(2273 \gamma ^6-1851 \gamma ^4-12957 \gamma ^2-2057\right)}{2 (\gamma^{2} -1)^4 }\cr
c_{13}(\gamma) =&-\frac{\gamma  \left(1575 \gamma ^6+1920 \gamma ^4-5177 \gamma ^2-1182\right)}{2 \left(\gamma ^2-1\right)^{5/2}}\cr
c_{14}(\gamma) =&-\frac{1249 \gamma ^6-1083 \gamma ^4-1053 \gamma ^2-9}{\left(\gamma ^2-1\right)^{5/2}}\cr
c_{15}(\gamma) =&-\frac{3 \left(225 \gamma ^6+600 \gamma ^5-315 \gamma ^4-1200 \gamma ^3+99 \gamma ^2+56 \gamma -9\right)}{\left(\gamma ^2-1\right)^{5/2}}\cr
c_{16}(\gamma) =&\frac{\gamma  \left(2100 \gamma ^6+1755 \gamma ^4-6422 \gamma ^2-1209\right)}{4 \left(\gamma ^2-1\right)^{5/2}}\cr
c_{17}(\gamma) =&\frac{9 \gamma  \left(2 \gamma ^2-3\right) \left(5 \gamma ^2-1\right)^2}{(\gamma^{2} -1)^3 }\cr
c_{18}(\gamma) =&\frac{1823 \gamma ^6-1221 \gamma ^4-13155 \gamma ^2-2039}{\left(\gamma ^2-1\right)^{5/2}}\cr
c_{19}(\gamma) =&\frac{\gamma  \left(2 \gamma ^2-3\right) \left(799 \gamma ^6-453 \gamma ^4-12003 \gamma ^2-2039\right)}
{(\gamma^{2} -1)^4 }\cr
c_{20}(\gamma) =&\frac{768 \left(8 \gamma ^6+14 \gamma ^4-116 \gamma ^2-19\right)}{\left(\gamma ^2-1\right)^{5/2}}\cr
c_{21}(\gamma) =&-\frac{3 \gamma ^2 \left(2 \gamma ^2-3\right) \left(175 \gamma ^6-355 \gamma ^4+185 \gamma ^2-37\right)}{4 (\gamma^{2} -1)^4 }\cr
c_{22}(\gamma) =&-\frac{4 \left(1823 \gamma ^6+6459 \gamma ^4-38115 \gamma ^2-6263\right)}{\left(\gamma ^2-1\right)^{5/2}}\cr
c_{23}(\gamma) =&-\frac{3871 \gamma ^6-2817 \gamma ^4-3747 \gamma ^2-75}{2 \left(\gamma ^2-1\right)^{5/2}}\cr
c_{24}(\gamma) =&-\frac{3 \gamma  \left(175 \gamma ^6-1255 \gamma ^4+1985 \gamma ^2-121\right)}{4 \left(\gamma ^2-1\right)^{5/2}}\cr
c_{25}(\gamma) =&\frac{3421 \gamma ^6-2067 \gamma ^4-5865 \gamma ^2+111}{2 \left(\gamma ^2-1\right)^{5/2}}\cr
c_{26}(\gamma) =&\frac{48 \gamma  \left(75 \gamma ^4-150 \gamma ^2+7\right)}{\left(\gamma ^2-1\right)^{5/2}}\cr
c_{27}(\gamma) =&-\frac{2 \left(4321 \gamma ^6+11973 \gamma ^4-75933 \gamma ^2-12553\right)}{\left(\gamma ^2-1\right)^{5/2}}\cr
c_{28}(\gamma) =&-\frac{64 \left(160 \gamma ^{12}-600 \gamma ^{10}-84 \gamma ^8+2058 \gamma ^6-1665 \gamma ^4-72 \gamma ^2+32\right)}{3 \left(\gamma ^2-1\right)^{11/2}}\cr
c_{29}(\gamma) =&-\frac{\gamma  \left(2 \gamma ^2-3\right) \left(1823 \gamma ^6-1221 \gamma ^4-13155 \gamma ^2-2039\right)}{(\gamma^{2} -1)^4 }\cr
c_{30}(\gamma) =&\frac{3150 \gamma ^7+1846 \gamma ^6-5775 \gamma ^5+198 \gamma ^4+5488 \gamma ^3-7518 \gamma ^2-1935 \gamma +258}{8 \left(\gamma ^2-1\right)^{5/2}}\cr
c_{31}(\gamma) =&\frac{1050 \gamma ^6+1696 \gamma ^5+389 \gamma ^4-1691 \gamma ^3-2241 \gamma ^2-1041 \gamma -114}{8 (\gamma -1) (\gamma ^2-1)^{3/2}}\cr
}}
\\
		\hline
	\end{tabular}
    \caption{Coefficient polynomials of the 5PM-1SF scattering angle $\theta^{(5,1)}
  =
  \sum_{k=1}^{31}
  c_k(\gamma) f_k(\gamma)$ of \eqn{thetadef}.
    }
	\label{table:Cpols}
\end{table*}

\clearpage
\newpage

\end{widetext}

\bibliographystyle{JHEP}
\bibliography{5pm-conservative}

\end{document}